# Safe Data Sharing and Data Dissemination on Smart Devices

Luc Bouganim, Cosmin Cremarenco, François Dang Ngoc,
Nicolas Dieu, Philippe Pucheral

I.N.R.I.A. Rocquencourt
78153 Le Chesnay Cedex, France
+33 1 39 63 52 50
<Firstname.Lastname>@inria.fr

## 1. INTRODUCTION

The erosion of trust put in traditional database servers and in Database Service Providers (DSP), the growing interest for different forms of data dissemination and the concern for protecting children from suspicious Internet content are different factors that lead to move the access control from servers to clients. Due to the intrinsic untrustworthiness of client devices, client-based access control solutions rely on data encryption. The data are kept encrypted at the server and a client is granted access to subparts of them according to the decryption keys in its possession. Several variations of this basic model have been proposed (e.g., [1, 6]) but they have in common to minimize the trust required on the client at the cost of a static way of sharing data. Indeed, whatever the granularity of sharing, the dataset is split in subsets reflecting a current sharing situation, each encrypted with a different key. Once the dataset is encrypted, changes in the access control rules definition may impact the subset boundaries, hence incurring a partial re-encryption of the dataset and a potential redistribution of keys.

Unfortunately, there are many situations where access control rules are user specific, dynamic and then difficult to predict. Let us consider a community of users (family, friends, research team) sharing data via a DSP or in a peer-to-peer fashion, it is likely that the sharing policies change as the initial situation evolves (relationship between users, new partners, new projects with diverging interest, etc.). Again, while the exchange of medical information is traditionally ruled by predefined sharing policies, these rules may suffer exceptions in particular situations (e.g., in case of emergency) [5] and may evolve over time (e.g., depending on the patient's treatment). Regarding parental control, neither Web site nor Internet Service Provider can predict the diversity of access control rules that parents with different sensibility are willing to enforce.

In the meantime, software and hardware architectures are rapidly evolving to integrate elements of trust in client devices. Secure tokens and smart cards plugged or embedded into different devices are exploited in a growing variety of applications (e.g., authentication, healthcare, digital right management). Thus, Secure Operating Environments (SOE) becomes a reality on client devices [10]. Hardware SOE guarantee a high tamper-resistance, generally on limited resources (tiny secured working and stable memories).

In [2], we exploited these new elements of trust in order to devise smarter client-based access control managers. The goal pursued is being able to evaluate dynamic and personalized access control rules on a ciphered input document, with the benefit of dissociating access rights from encryption. The considered input documents are XML documents, the de-facto standard for data exchange. We proposed a streaming evaluator of access control rules based on non-deterministic automata and designed a streaming index structure allowing skipping the irrelevant parts of the input document. To demonstrate the effectiveness of the approach, we made experiments considering a smart card as the target SOE. Given the current smart card hardware limitations, we relied on a prototype written in C and running on a cycle accurate smart card hardware simulator.

The objective of this demonstration is threefold:

1. To validate the solution proposed in [2] on a real smart card platform, including the non-deterministic automata engine and the skip index structure.
2. To show how our smart card engine can be integrated in a distributed architecture including the smart card, the server and the user terminal. Indeed, the tamper resistance of the access control relies not only on the SOE but also on the whole environment (e.g., communication protocol, access rights update protocol, etc.). These important issues have not been deeply discussed in [2], where the focus was put on the components running inside the SOE.
3. To illustrate the generality of the approach and the easiness of deployment of the proposed infrastructure through two very different applications.

This paper is organized as follows. Section 2 recalls from [2] the foundation of our proposal required to weight up the value of the demonstration. Section 3 presents the demonstration platform and the way we plan to validate our techniques.

## 2. DESIGN PRINCIPLES

To illustrate the approach, let us consider different users willing to exchange safely sensitive information (i.e., XML documents in our context) located in an untrusted data store (e.g., a DSP).

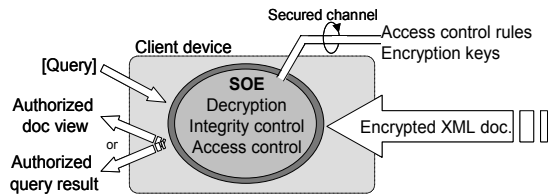
**Figure 1. Target architecture**

Basically, when a user issues a query to the data store, the SOE fetches the appropriate encrypted XML document from the server, decrypts it, checks that it has not been tampered, enforces the access control on it and delivers the authorized subpart matching the query. While this example illustrates a pull-based scenario, our approach can support push-based scenarios (e.g., selective data dissemination) in a very similar way. In the following, we first depict the abstract architecture; then we present the access control model and finally describe the proposed streaming access control mechanism.

## 2.1 Target architecture

Figure 1 pictures an abstract representation of the target architecture for the applications mentioned in the introduction. The access control being evaluated on the client, the client device has to be made tamper resistant thanks to a Secure Operating Environment (SOE) (e.g., a smart card). We make the following traditional assumptions on the SOE: 1) the code executed by the SOE cannot be corrupted, 2) the SOE has at least a small quantity of secure stable storage (to store secrets like encryption keys), 3) the SOE has at least a small quantity of secure working memory (to protect sensitive data structures at processing time).

In our context, the SOE is in charge of decrypting the input document, checking its integrity and evaluating the access control policy corresponding to a given (document, subject) pair. Under the assumption that the SOE is secure, the only way to mislead the access control rule evaluator is to tamper the input document, for example by substituting or modifying encrypted blocks, thus motivating the encryption and integrity checking.

Access control policies as well as the key(s) required to decrypt the document can be either permanently hosted by the SOE, refreshed or downloaded via a secure channel from different sources (trusted server, license provider, server managing parent's consent, etc).

## 2.2 Access Control Model

Several authorization models have been recently proposed for regulating access to XML documents. We introduce below a simplified access control model for XML, inspired by Bertino's model [1] and Samarati's model [3] that roughly share the same foundation. Subtleties of these models are ignored for the sake of simplicity. In this simplified model, access control rules, or access rules for short, take the form of a 3-uple <*sign*, *subject*, *object*>.

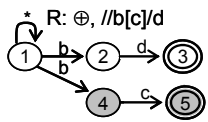

**Figure 2. Access control rule automaton**

*Sign* denotes either a permission (positive rule) or a prohibition (negative rule) for the read operation. *Subject* is self-explanatory. *Object* corresponds to elements or subtrees in the XML document, identified by an XPath expression. We consider here a rather robust subset of XPath denoted by $XP^{\{[],*,//\}}$ [7]. This subset, widely used in practice, consists of node tests, the child axis (/), the descendant axis (//), wildcards (*) and predicates or branches [...]. The cascading propagation of rules is implicit in the model, meaning that a rule propagates from an object to all its descendants in the XML hierarchy. Due to this propagation mechanism and to the multiplicity of rules for a same user, a conflict resolution principle is required. Conflicts are resolved using two policies: 1) *Denial-Takes-Precedence,* which states that if two rules of opposite signs apply on the same object, then the negative one prevails and 2) *Most-Specific-Object-Takes-Precedence,* which states that a rule which applies directly to an object takes precedence over a propagated rule.

## 2.3 Efficient streaming access control

The streaming requirement is twofold. First, the evaluator must adapt to the memory constraint of the SOE, thereby precluding materialization (e.g., building a DOM representation of the document). Second, some target applications mentioned in the introduction are likely to consume streaming documents (e.g., selective data dissemination). Efficiency is, as usual, an important concern, and leads to rely on indexing to quickly converge towards the authorized parts of the input document, while skipping the others. Indexing is of utmost importance considering the two limiting factors of the target architecture: the cost of decryption in the SOE and the cost of communication between the SOE, the client and the server.

**Access rules evaluation**
At first glance, streaming access control resembles the well-known problem of XPath processing on streaming documents [8], which has been widely studied notably in the context of XML filtering [4]. While access rules are expressed in XPath, the nature of our problem differs significantly from the preceding ones because rules are not independent, bringing two new issues: 1) at parsing time the evaluator must be capable of determining the set of rules targeting a given node and deciding which one apply according to the conflict resolution policies and 2) some rules may be inhibited by others according to the conflict resolution policies, thereby optimizations such as suspending evaluations of rules can be devised.

As streaming documents are considered, we make the assumption that the evaluator is fed by an event-based parser (e.g., SAX) raising *open*, *value* and *close* events respectively for each opening, text and closing tag in the input document. Each access rule is represented by a non-deterministic automaton, as pictured in Figure 2. This automaton is made up of a *navigational path* (in white in the figure) representing the XPath without its predicate and *predicate paths* (in gray in the figure) appended to it.

Basically, when an open or a value event is received, all the automata are checked and go to their next state. Upon receiving a close event, all the automata backtrack. To manage these automata efficiently, we use a stack that keeps track of active states, materializing all the possible paths that can be followed on the non-deterministic automata. In order for a rule to apply, all its final states must be reached. This is controlled using a predicate set which records all the final states of predicates that have been reached. In some cases, the final state of a navigational path may be reached while those of its predicate paths are not. In these cases, the rule is said to be pending, meaning that the nodes upon which it applies are to be delivered only if, later on in the parsing, all the predicate paths are found to reach their final states. Finally, propagation of rules as well as conflicts are managed with a sign stack which keeps on the top, the current sign that is propagated if no other rule applies. More details can be found in [2].

**Skip index**

To reduce the flow of data received by the SOE and thus the decryption time, we devise a new indexation structure that enables to skip irrelevant (i.e., forbidden) parts of the documents. The first distinguishing feature of the required index is the necessity to keep it encrypted outside of the SOE to guarantee the absence of information disclosure. The second distinguishing feature (related to the first one and to the SOE storage capacity) is that the SOE must manage the index in a streaming fashion, similarly to the document itself. These two features lead to design a very compact index (its decryption and transmission overhead must not exceed its own benefit), embedded in the document in a way compatible with streaming. For these reasons, we concentrate on indexing the structure of the document, pushing aside the indexation of its content. The objective of the index is to detect rules and queries that cannot apply inside a given subtree, with the expected benefit to skip this subtree if it turns out to be forbidden or irrelevant wrt the query. Keeping the compactness requirement in mind, the minimal information required to achieve this goal is the set of element tags that appear in each subtree (to check whether an access rule automaton is likely to reach its final state) as well as the subtree size (to make the skip actually possible). Although this metadata does not capture the tag nesting, it reveals oneself as a very effective way to filter out irrelevant access control rules. For ensuring compactness, we compress the document structure using a dictionary of tags [9] and encode the set of tags thanks to a bit array referring to the tag dictionary. To further reduce the indexing overhead, we apply recursive compression on both the set of tags bit array and the subtree size. More details can be found in [2].

## 3. THE DEMONSTRATOR

The demonstration platform includes the whole prototype architecture as well as two applications. The architecture, depicted in Figure 3, is composed of:

- a terminal connected to the smart card. It contains a proxy allowing the applications to communicate easily with the different elements of the architecture through an XML API independent of the underlying protocols (JDBC, APDU[1]).
- a DSP which hosts encrypted XML documents shared by users as well as encrypted access rules. Both are encrypted using secret keys exchanged between users thanks to a public key infrastructure (PKI)[2].
- a smart card in charge of enforcing access control. For the demonstration, we use e-gate smart cards developed by Axalto,

---

[1] Application Protocol Data Unit: Communication protocol between the terminal and the smart card
[2] In the demonstration, we will not use a PKI infrastructure but rather simulate it to keep the demonstration independent of a network connection. Moreover, PKI is a well-known technique that need not be demonstrated.

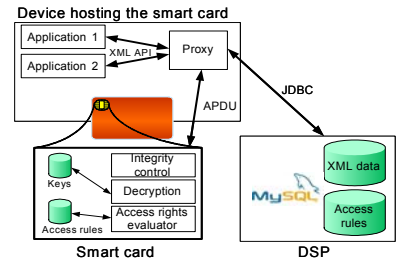

**Figure 3. Architecture**

a subsidiary of Schlumberger. These smart cards provide a powerful CPU and strong security features but still have a limited memory (only 1 KB of RAM available for on-board applications) and a low bandwidth (2KB/s).

The advantage provided by this architecture is twofold. First, the complexity of the access control, query and security management is confined in the smart card and its proxy, so that the application developer can concentrate on the application logic. Second, the proposed architecture complies with the standards. Documents are described in XML and both access control rules and queries are expressed in XPath, two very popular W3C standards.

To assess the generality of the approach, the features of our prototype will be demonstrated through two applications. The first application deals with collaborative works among a community of users while the second one deals with the selective dissemination of multimedia streams through unsecured channels. They are representative of two rather different profiles regarding the way the information is accessed (pull vs. push), the type of this information (textual vs. video) and the response time requirements (user patience / real time).